\documentclass[twocolumn,showpacs,preprintnumbers,amsmath,amssymb]{revtex4}

\usepackage{graphicx}
\usepackage{dcolumn}
\usepackage{bm}

\begin{document}


\title{Image-potential induced states at metal surfaces}
\author{P. M. Echenique$^{1,2}$, J. M. Pitarke$^{2,3}$, E. V. Chulkov$^{1,2}$,
and V. M. Silkin$^2$}
\affiliation{$^1$Materialen Fisika Saila, Kimika Fakultatea,
Euskal Herriko Unibertsitatea,
1072 Posta kutxatila, E-20080 Donostia, Basque Country\\
$^2$Donostia International Physics Center (DIPC) and Centro Mixto
CSIC-UPV/EHU,
Donostia, Basque Country\\
$^3$ Materia Kondentsatuaren Fisika Saila, Zientzi Fakultatea,
Euskal Herriko Unibertsitatea,\\ 644 Posta kutxatila, E-48080 Bilbo, Basque
Country}
\date\today

\begin{abstract}
The unequivocal observation in the early 80's of a new class of unoccupied 
surface states that are bound to the vacuum level of a variety of
metal surfaces has led over the last two decades to an active area of
research in condensed-matter and surface physics. Here we discuss 
the key ingredients of the theory of these so-called image states,
which are originated in the long range nature of the image potential
outside a solid. 
\end{abstract}

\maketitle

\section{Introduction}

It was shown by Cole and Cohen \cite{cole} and by Shikin \cite{shikin} 
that for negative electron affinity
materials the image potential outside the surface should trap electrons in
Rydbergs states, thereby explaining the experimentally observed trapping of
electrons at the surface of liquid helium \cite{sommer,grimes1,grimes2}.
These so-called
image states were then invoked \cite{echenique1,rundgren} to explain the fine 
structure that had been observed in the difraction of
low-energy electrons from metal surfaces with a band gap near the vacuum level
\cite{mcrae1,lauzier,mcrae2}. Echenique and Pendry \cite{echenique2}
also investigated the
observability of image states via low-energy-electron difraction (LEED) 
experiments, and a discussion
of the lifetime broadening of these states led them to the important
conclusion that they could in principle be resolved for all
members of the Rydberg series.

Another type of surface states in the band gaps of free-electron-like
$s$,$p$ bands, which can occur even for a step barrier in the absence
of the image potential, were predicted by Shockley \cite{shockley}.
These states have been observed with the use of
photoemission in a variety of metal surfaces
\cite{gartland,heimann,levinson,kevan,hulbert86}.

In angle-resolved photoemission experiments a sample is illuminated with
monochromatised photons and the energy spectrum of the photoemitted
electrons is examined, thus providing information about the density of
occupied states. In order to investigate unoccupied states in the range
between the Fermi and the vacuum levels, one has to use other methods
such as inverse photoemission \cite{pendry}. In these experiments 
electrons with a given energy are incident on a solid surface
and the energy of the emitted photon is measured.
Hence, this technique allows to map out the features
of the unoccupied density of states and to derive the energy and momentum
of bound states by simply measuring the energy and momentum of the incident 
electron and the
energy of the emitted photon.

Johnson and Smith \cite{johnson1} pointed out that image-potential induced 
bound states were 
potentially observable
by angle-resolved inverse photoemission. By using this technique, Dose {\it
et al.} \cite{dose1} and Straub and Himpsel \cite{himpsel1} reported the first
conclusive experimental evidence for image-potential bound states at the
(100) surfaces of copper and gold. Since then, several observations of
image states were made using this technique
\cite{johnson2,reihl,johnson3,himpsel2,smithr,himpsel3,donath}, 
and also the new high-resolution techniques 
of two-photon photoemission (TPPE) \cite{himpsel4,haight,fauster} 
and time-resolved two-photon photoemission
(TR-TPPE) \cite{schoenlein1,schoenlein2}. In TPPE, 
intense laser radiation is used to populate an unoccupied
state with the first photon and to photoionize from this intermediate
state with the second photon. In TR-TPPE, the probe pulse which ionizes the
intermediate state is delayed with respect to the pump pulse which
populates it, thus providing a direct measurement of the intermediate
state lifetime \cite{hertel,wolf,merry,harris,hofer,knoesel,shumay}. 
Since image states are typically decoupled from 
bulk states, their long lifetimes make them to be ideally suited as 
intermediate states. Both image states and the image-like form of the
surface barrier outside a solid also play a role in scanning tunneling
microscopy (STM) \cite{binnig,golovchenko,pitarke1,pitarke2}.

The binding energies of image sates were originally investigated in terms
of a simple multiple-scattering approach\cite{gurman}, in which surface 
states
are viewed as waves that are trapped and multiply scattered between the bulk 
crystal on one side and the
surface barrier on the other \cite{mcrae2,echenique2}. 
Soon after the
first observation of these states, it was shown that the crystal-face
dependent binding energies of both Shockley and image states could be
understood in terms of the multiple-scattering approach \cite{johnson4}, 
and this
approach was then combined with a nearly-free-electron (NFE) description
of the bulk band structure \cite{goodwin} to predict Shockley and 
image-state binding
energies that were in reasonable agreement with photoemission and inverse
photoemission measurements \cite{smith85,garrett,ortuno,smith89}. Wave-function
matching techniques were also applied to obtain both binding energies and wave
functions of surface states and to demonstrate that the spatial extent
of these states is determined by their binding energy \cite{johnson5}.

Since the pioneering work of Appelbaum and Hamann \cite{appelbaum}, 
first-principles
calculations of the electronic structure of solid surfaces have been
carried out within the framework of density-functional theory (DFT) \cite{kohn1} 
by replacing the full many-body Schr\"odinger equation by an effective 
one-particle equation that is typically solved self-consistently
in the so-called local-density approximation (LDA) \cite{kohn2}. 
Although this approximation
ceases to be valid in the tail of the electron density, where the asymptotic
value of the one-particle LDA effective potential does not reproduce
the correct image-like behaviour, LDA calculations have proved
to be succesful in the description of both bulk states travelling away
from the surface and Shockley surface states decaying exponentially
away from the surface. Nevertheless, image states can only be described
within a correct non-local description of the one-particle potential
in the vacuum side of the surface. By matching a long-range image potential
onto the self-consistent LDA crystal potential, the experimentally
observed binding energies and effective masses of image states on a
variety of clean metal surfaces and on MgB$_2$(0001) were reproduced
\cite{johnson6,nekovee1,nekovee2,chulkov00,crampin1,gao1,silkin99,silkin01}. 
In a more sophisticated
approach, Eguiluz {\it et al.} \cite{eguiluz1} obtained the correct asymptotic 
behaviour
of the exchange-correlation potential at large distances outside a
jellium surface by solving the so-called Sham-Sch\"ulter integral equation,
which relates the exact exchange-correlation potential of DFT
to the electron self-energy of many-body theory \cite{sham1}. 
A parametrization of this non-local
potential was then used to obtain a Rydberg series of image states in the
(100) and (111) surfaces of Al and Pd \cite{eguiluz2}.    

Alternatively, the excitation energies and quasiparticle wavefunctions of 
unoccupied states can be calculated rigorously from a one-particle
Schr\"odinger equation containing an energy-dependent non-local complex
self-energy \cite{hedin}. Simplified free-electron gas (FEG) models
of the electron self-energy were employed by Echenique and collaborators
\cite{echenique3,echenique4,bausells1,bausells2} to calculate both binding 
energies
and a local effective potential.
Full GW \cite{gunnarson1} self-energy calculations of the
binding energy and the local effective potential were reported in Refs.
\cite{deisz1,deisz2} for the lowest-lying 
image resonance at a
jellium surface, showing that the effective potential nearly coincides
with the exchange-correlation (xc) potential of DFT obtained in Ref. \cite{eguiluz1}.
Furthermore, it was shown in Refs. \cite{deisz1,deisz2}
that upon inclusion of long-range correlations into the xc
potential \cite{eguiluz1} the DFT eigenfunctions and eigenvalues for
a jellium surface are extremely good approximations to their quasiparticle
counterparts. More recently, full GW self-energy calculations for quasiparticle
states at the (111) surface of real Al where reported \cite{godby1},
within $1.5\,{\rm eV}$ of the vacuum energy, which led to a
first-principles evaluation of the image-plane position for this surface.
The image plane for the many-electron system was found to be closer to
the surface than that for the classical response to external fields of
charges \cite{lang2}, and it was also found to be significantly modified
by the atomic structure of the surface.

The finite lifetime of excited states 
can be obtained from the knowledge of the imaginary part of the electron
self-energy \cite{chemphys}. As the center of gravity of image states 
is located outside
the solid, Echenique and Pendry \cite{echenique2} assumed that the broadening 
of these states
is dictated by their small overlap with the bulk region of the surface;
for a constant value of the imaginary part of the bulk self-energy,
they found that the image-state broadening is
for all the levels in the Rydberg series smaller than the level spacing,
thereby demonstrating the integrity of image states at metal surfaces.
By combining TR-TPPE with the coherent excitation of several quantum states
six image states were recently resolved on a (100) surface of 
Cu \cite{hofer}, and their
linewidths were found to approximately increase with the quantum number
as originaly predicted in Ref. \cite{echenique2}.

The first quantitative evaluation of image-state lifetimes was reported
in Ref. \cite{echenique5}. In this calculation, the image-state wave
functions were approximated by hydrogenic-like wave functions with no
penetration into the solid and a simplified FEG model was used to approximate
the electron self-energy. In subsequent calculations the penetration
of the image-state wave function into the crystal was allowed
\cite{echenique6,echenique7},
and the role that the unoccupied part of the narrow Shockley surface state 
on the (111) surfaces of Cu and Ni plays in the decay of the $n=1$ image state
on these surfaces was investigated by Gao and Lundqvist \cite{gao2}. 
In this work,
the image-state wave functions were also approximated by hydrogenic-like
wave functions with no penetration into the solid, a simplified
parametrized form was used for the Shockley surface-state wave function, and
screening effects were neglected altogether. A GW calculation
of the imaginary part of the electron-self energy near a jellium surface
was also reported \cite{deisz3}, showing the key role that a full evaluation 
of this quantity may play in the description of surface-state lifetimes. 

The first self-consistent many-body calculations of image-state lifetimes on
noble and simple metals were reported only recently \cite{chulkov1,chulkov2},
and good agreement with experimentally determined decay 
times \cite{hofer,knoesel,shumay} was found. In these calculations
all wave functions and energies were obtained by solving a one-particle
Schr\"odinger equation with a realistic one-dimensional model potential,
the potential variation in the plane parallel to the surface was considered
through the introduction of an effective mass, and the electron
self-energy was evaluated in the GW approximation. Later, self-consistent 
calculations
of the key role that the partially unoccupied Schokley surface state
plays in the decay of image states on Cu(111) were carried out \cite{osma},
and the inclusion of short-range xc effects was also investigated \cite{sarria}.
It was demonstrated that although the presence of short-range
exchange and correlation between screening electrons significantly enhances
the decay probability of image states, this enhancement happens to be more 
than compensated by the large reduction in the decay rate produced by the
presence of a xc hole around the image-state electron itself.

Recently, photohole lifetimes of Shockley surface states were investigated 
in a variety of metal
surfaces with the use of high-resolution angle-resolved photoemission
spectroscopy \cite{McDougall,goldmann,balasu,valla,reinert,balu01}. 
Scanning tunneling
spectroscopy was also used to determine the lifetime of excited
holes at the edge of the partially occupied Shockley surface-state band on the
(111) face of noble metals \cite{stm1}, and to measure the lifetime
of Shockley surface-state and surface-resonance electrons as a function
of their energy \cite{stm2}. GW calculations \cite{surface1,chusiss,surface2}
demonstrated that the decay of Shockley surface-state holes is dominated
by two-dimensional electron-electron interactions screened by the 
underlying three-dimensional electron system, and electron-electron
interactions were then combined with the electron-phonon coupling to find
good agreement with the experiment \cite{surface1}. These theoretical investigations
were extended to study the dynamics of surface-state and surface-resonance
unnocupied electronic states in Cu(111) \cite{surface3}, showing that,
contrary to the case of surface-state holes, relatively major contributions to the
e-e interaction of surface-state electrons above the Fermi level
come from the underlying bulk electrons, and thereby giving an interpretation
to the measurements reported in Ref. \cite{stm2}.

Unless otherwise is stated, atomic units are used throughout this paper, i.e.,
$e^2=\hbar=m_e=1$. The atomic unit of length is the Bohr radius,
$a_0=\hbar^2/m_e^2=0.529{\textrm\AA}$, the atomic unit of
energy is the Hartree, $1\,{\rm Hartree}=e^2/a_0=27.2\,{\rm eV}$, and
the atomic unit of velocity is the Bohr velocity, $v_0=\alpha\,
c=2.19\times 10^8{\rm cm\,s^{-1}}$,
$\alpha$ and $c$ being the fine structure constant and the velocity of
light, respectively.

\section{Basics of image states}

Image states are quantum states trapped in the image-potential well outside
the surface of a material with negative electron affinity. In the case of
a hard-wall substrate that occupies the half-space $z<0$ and has a static
dielectric constant $\epsilon$, the asymptotic form of the potential
experienced by an electron in the half space $z>0$ is the classical
image potential
\begin{equation}\label{image}
V(z)=-\frac{1}{4z}\,\frac{\epsilon-1}{\epsilon+1}.
\end{equation}
If the substrate is assumed to be infinitely repulsive, then the electron
wave functions and energies are of the form
\begin{equation}\label{psi}
\Psi({\bf r})=\frac{1}{\sqrt A}\,\psi(z)\,{\rm e}^{{\rm i}\,{\bf k}_\parallel 
\cdot{\bf r}_\parallel}
\end{equation}
and
\begin{equation}\label{energy}
E=\varepsilon+\frac{1}{2}\,{\bf k}_\parallel^2,
\end{equation}
where ${\bf k}_\parallel$ and ${\bf r}_\parallel$ are wave and position 
vectors parallel to the surface, $A$ is the
normalization area, and $\psi(z)$ and $\varepsilon$ are the eigenfunctions
and eigenvalues of the one-electron Schr\"odinger equation
\begin{equation}\label{schrodinger}
\frac{d^2}{dz^2}\,\psi(z)+V(z)=\varepsilon\,\psi(z).
\end{equation}
Introduction of Eq. (\ref{image}) into Eq. (\ref{schrodinger}) gives
rise to a Rydberg series of image-potential induced bound states (see Fig. 1) 
with
\begin{equation}\label{rydberg}
\psi_n(z)\propto z\,\psi_n^{\rm hydrogen}(z/4)~~,~~~~ n=1,2,\dots
\end{equation}
and
\begin{equation}\label{rydberg2}
\varepsilon_n=-\frac{1}{32\, n^2}\,\frac{\epsilon-1}
{\epsilon+1}~~,~~~~ n=1,2,\dots,
\end{equation}
where $\psi_n^{\rm hydrogen}(z)$ represent the well-known wave functions of
all possible $s$-like ($l=0$) bound states of the hydrogen atom. Hence, the
characteristic 'Bohr radius' of image states is expected to be
\begin{equation}
a_B=\frac{4}{n^2}\,\frac{\epsilon-1}
{\epsilon+1}\,a_0~~,~~~~ n=1,2,\dots.
\end{equation}
In the case of
liquid helium, the static dielectric constant is very close to unity
($\epsilon=1.0572$ for $^4$He \cite{smee}), the binding energy of 
image states is therefore
of the order of a few tenths of meV 
($\varepsilon=-0.66/n^2\,{\rm meV}$), much smaller than the actual barrier
height \cite{woolf}, and the assumption of an infinitely repulsive 
substrate seems reasonable.

\begin{figure}
\includegraphics[width=0.4\textwidth,height=0.6\textwidth]{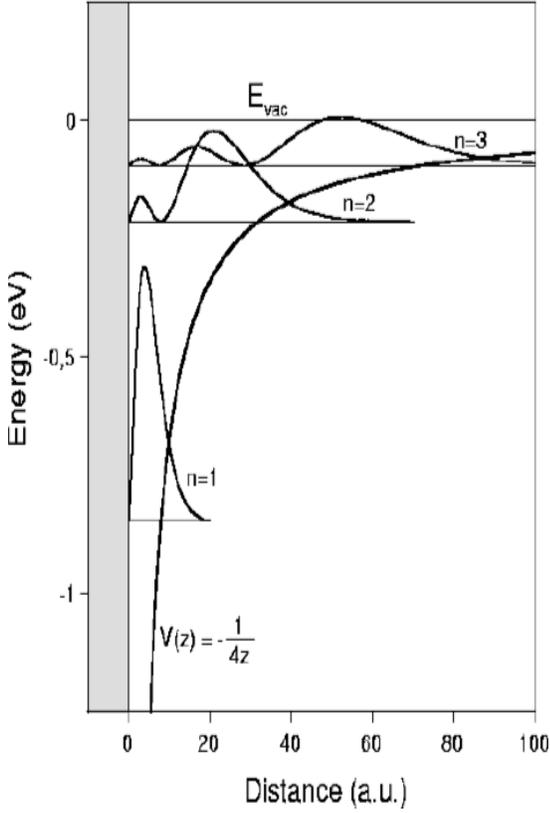}
\caption{ Image potential and first image states of a Rydberg series plotted
together with the infinite crystal barrier. }  
\label{fig1}
\end{figure} 

In most metals there is no repulsive barrier for an electron
to return to the metal, but in a few metals there is a band gap near the
vacuum level which may trap an electron in the quantum states of
the image potential (see Fig. 2). For metals the static dielectric constant 
is infinitely negative ($\epsilon\to -\infty$), and image-state
binding energies are of the order of a few tenths of eV
($\varepsilon=-0.85/n^2\,{\rm eV}$). As these energies are comparable
with the typical width of the band gap, the condition of an infinitely
repulsive substrate must be removed. A simple way to overcome this
dificulty can be found in the multiple scattering approach developed
in Ref. \cite{echenique2}, which pictures surface states
as standing waves of an electron bouncing back and forth between the
bulk crystal and the surface barrier.

\begin{figure}
\includegraphics[width=0.4\textwidth,height=0.6\textwidth]{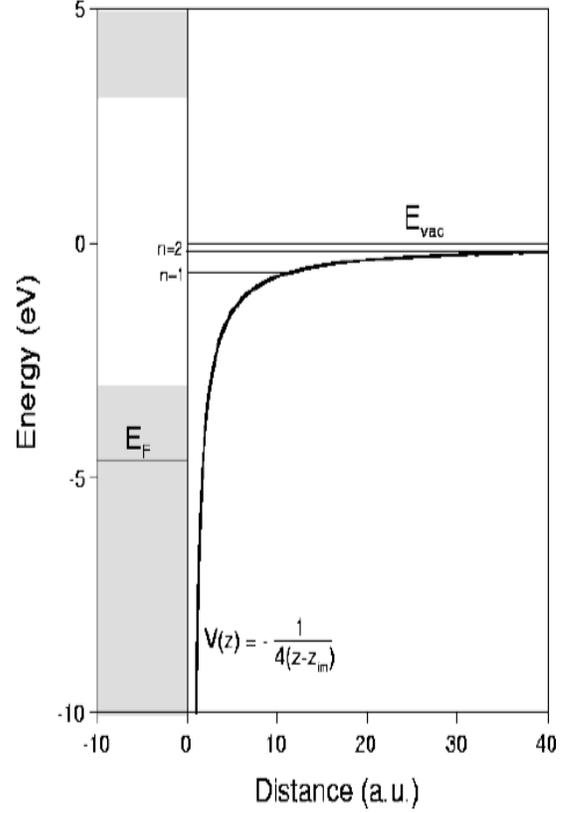}
\caption{Image potential for realistic negative affinity materials.}
\label{fig2}
\end{figure}                                                                    

\subsection{Multiple scattering approach}

Consider an electron wave $\psi_+$ propagating freely from the crystal 
termination
at $z=0$, which we choose to be located at one-half the atomic spacing beyond
the last atomic layer, to the surface barrier at $z=z_{\rm im}+z_c$ ($z_{\rm im}$ is the
image-plane position), 
as shown in Fig. 3. This wave will be Bragg reflected both at the
crystal ($z=0$) and at the surface ($z=z_{\rm im}+z_c$) barrier. 
After both reflexions, the total reflectivity will be
\begin{equation}
R=r_c\,r_b\,{\rm e}^{{\rm i}\,\left[\phi_B+2\,\kappa\,(z_{\rm im}+z_c)+\phi_C\right]},
\end{equation}
where $\phi_B$ and $\phi_C$ denote the phase changes between incident
and reflected waves that are produced by the crystal and the surface
barriers, respectively, and $\kappa$ is the perpendicular component of 
the free-electron wave vector at $0<z<z_{\rm im}+z_c$. Summing the repeated 
scatterings gives
\begin{equation}\label{amplitude}
\psi_+\propto \left[1-r_B\,r_C\,{\rm e}^{{\rm i}\,
\left[\phi_c+2\,\kappa\,(z_{\rm im}+z_c)+\phi_b\right]}\right]^{-1}.
\end{equation}
Bound states correspond to the poles of the total amplitude dictated by Eq.
(\ref{amplitude}), and occur when 
\begin{equation}
r_B=r_C=1
\end{equation}
 and
\begin{equation}\label{bohr}
\phi_B+2\,k\,(z_{\rm im}+z_c)+\phi_C=2\pi\,n~~,~~~~ n=0,1,2,\dots.
\end{equation}  

\begin{figure}
\includegraphics[width=0.4\textwidth,height=0.6\textwidth]{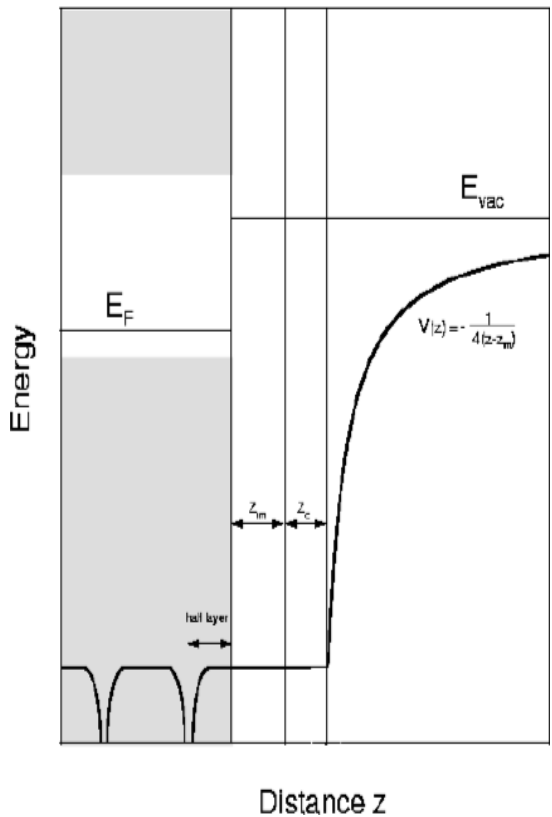}
\caption{Schematic representation of the full one-electron potential
together with the image potential and image plane position $z_{im}$.}
\label{fig3}
\end{figure}

The phase change produced by any general potential barrier is obtained
by matching the wave function $\psi(z)$ and its derivative $\dot\psi(z)$ 
at the boundary of the
barrier with the total (incident plus reflected) wave. One finds
\begin{equation}\label{phase}
\phi={\rm arg}\,\frac{{\rm i}\,\kappa-L}{{\rm i}\,\kappa+L},
\end{equation}
where $L$ is the logarithmic derivative $L=\dot\psi/\psi$ of the wave function 
at the boundary
of the barrier.

\subsubsection{Infinite crystal barrier}

In the limiting case of an infinite crystal barrier at $z=0$ and a pure 
image potential at $z>0$,
the crystal-induced phase is easily found to be $\phi_C=\pi$, $\kappa$ is
infinite for finite energies, $z_{\rm im}=z_c=0$, and the eigenfunctions
of the one-particle Schr\"odinger equation that are finite at infinite are the
Whittaker functions
\begin{equation}
\psi(z)=W_{\lambda,1/2}(z/2\lambda),
\end{equation}
where
\begin{equation}
\lambda=(-32\,\varepsilon)^{-1/2}.
\end{equation}
For $\lambda=n$ ($n=1,2,\dots$), $W_{\lambda,1/2}(0)=0$, $L$ is infinite, and
Eq. (\ref{phase}) yields $\phi_B=(2n-1)\,\pi$, while for $\lambda\neq n$ 
($n=1,2,\dots$)
Eq. (\ref{phase}) yields
$\phi_B=2\,n\,\pi$. Hence, in this limiting case $\phi_B$ is a step-like
function of $\varepsilon$ and the Bohr-like quantization condition
of Eq. (\ref{bohr}) is satisfied by the Rydberg series of bound states
dictated by Eq. (\ref{rydberg2}) with $\epsilon\to\infty$, i. e.
\begin{equation}
\varepsilon_n=-\frac{1}{32\, n^2}
~~,~~~~ n=1,2,\dots.
\end{equation}
The $n=0$ condition of Eq. (\ref{bohr}) is never satisfied.

\subsubsection{Abrupt step barrier}

Alternatively, one may consider the limiting case of an abrupt step barrier
of height $V_0$  at 
$z=z_c$. In this case,
\begin{equation}
\phi_B=2\,\arctan\,\left\{-\left[-\varepsilon/(\varepsilon-V_0)\right]^{-1/2}
\right\}
\end{equation}
and the $n=0$ condition ($\phi_B=-\phi_C$) can be satisfied,
which yields the well-known crystal-induced surface state that exists
in the presence of a Shockley-inverted gap. Image states are not present
in this model. 

\subsubsection{Quantum defect}

In the more general case of an image potential of the form 
\begin{equation}
V(z)=-\frac{1}{4(z-z_{\rm im})}
\end{equation}
that joins the inner potential $V_0$ at $z=z_c$, both the crystal-induced
$n=0$ Shockley state and the infinite Rydberg series of image states may exist.
While the $n=0$ solution occurs at energies that are close to the Fermi level,
the $n\neq 0$ solutions occur at energies that lie near the vacuum level. For
these energies, the variation of the crystal-induced phase $\phi_C$ is small 
and the variation of the surface-barrier induced phase $\phi_B$
can be described by the Wentzel-Kramers-Brillouin (WKB) expression\cite{mcrae0}
\begin{equation}\label{phib}
\phi_B=\left(2/\lambda-1\right)\,\pi.
\end{equation} 
Neglecting the energy dependence of $\phi_C$, substitution of 
Eq. (\ref{phib}) into Eq. (\ref{bohr}) yields
\begin{equation}
\varepsilon_n=-\frac{1}{32\, (n+a)^2}
~~,~~~~ n=1,2,\dots,
\end{equation}
where
\begin{equation}\label{defect}
a={1\over 2}\,\left(1-\phi_C/\pi\right).
\end{equation}

Here, the so-called quantum defect parameter $a$ is related to the
bulk band structure through the crystal-induced phase $\phi_C$. In the
limiting case of an infinite crystal barrier, $\phi_C=\pi$ and
the quantum defect parameter is $a=0$.

In order to take explicit account of the variation of $\phi_C$ with
energy and to describe the binding energies of both crystal-induced ($n=0$)
and image-potential induced ($n\neq 0$) surface states, Smith combined the 
multiple scattering description of these states with a NFE description
of the bulk band structure and the use of only two bands\cite{smith85}.

\section{Two-band model}

In the two-band approximation to the NFE band structure of a solid, electron
energies $E({\bf k})$ are the solutions of
\begin{eqnarray}\label{det}
\begin{vmatrix}
E({\bf k})-E^0({\bf k})&& -V_{\bf g}\\\\
-V_{\bf g}&&E({\bf k})-E^0({\bf k}-{\bf g})
\end{vmatrix},
\end{eqnarray} 
where $E^0({\bf k})={\bf k}^2/2$ is the free-electron energy and $V_{\bf g}$ 
is the pseudopotential coefficient associated
with the reciprocal lattice vector ${\bf g}$.
At the ${\bf k}={\bf g}/2$ point the degenerate free-electron levels
$E^0({\bf k})$ and $E^0(-{\bf k})$ are separated by a band gap of magnitude
$2|V_{\bf g}|$. Within this gap, solutions of Eq. (\ref{det}) with real
energy are still possible with a complex
$k_z=p+{\rm i}\,q$ corresponding to wave functions $\psi(z)$ that 
decay away from the surface into the bulk.

In the case of a gap that is opened by potential Fourier components
corresponding to ${\bf g}$ vectors that are normal to the surface,
Eq. (\ref{det}) yields
\begin{equation}
p=g_z/2,
\end{equation}
\begin{equation}
q^2/2=\sqrt{4\,(\varepsilon-V_0)\,E^0({\bf g}/2)+V_{\bf g}^2}-
\left[\varepsilon+E^0({\bf g}/2)-V_0\right],
\end{equation}
and
\begin{equation}
\psi(z)={\rm e}^{q\,(z-z_0)}\,\cos(p\,(z-z_0)+\delta).
\end{equation}
Here, $z_0$ is the $z$ coordinate of a surface atom, $V_0$ is
the crystal inner potential, and $\delta$ represents
a phase parameter,
\begin{equation}
\sin 2\delta=-p\, q/V_{\bf g},
\end{equation}
which in the presence of a Shockley-inverted band gap ($V_{\bf g}>0$
\cite{forstmann})
varies from $-\pi/2$ at the bottom
of the gap to $0$ at the top of the gap. Matching at $z=0$
to a wave function of the form
\begin{equation}
\psi(z)={\rm e}^{-{\rm i}\,\kappa\,z}+r_C\,{\rm e}^{{\rm i}\,\phi_C}\,
{\rm e}^{-{\rm i}\,\kappa\,z},
\end{equation}
one finds,
\begin{equation}
r_C=1
\end{equation}
and
\begin{equation}\label{phic}
\kappa\,\tan\left(\phi_C/2\right)=\frac{g}{2}\,\tan\left(\frac{\pi}{2}+
\delta\right).
\end{equation}
Hence, in the case of a Shockley-inverted band gap the phase $\phi_C$ varies from
0 at the bottom of the gap to $\pi$ at the top, and the quantum defect of Eq.
(\ref{defect}) varies from $a=1/2$ at the bottom to $a=0$ at the top. 

In the more general case of a band gap that is opened by potential
Fourier components corresponding to ${\bf g}$ vectors that are not
necessarily normal to the surface, and at the points of high symmetry
at the edge of the surface Brillouin zone where $k_x=g_x/2$ and
$k_y=g_y/2$, there are two branches $\phi_C^+$ and $\phi_C^-$
depending on whether the wave function $\phi(z)$ within the crystal
is of even or odd symmetry with respect to surface atoms,
\begin{equation}\label{phicp}
\kappa\,\tan\left(\phi_C^+/2\right)=\frac{g}{2}\,\tan\left(\frac{\pi}{2}+
\delta\right)
\end{equation}
and
\begin{equation}\label{phicm}
\kappa\,\tan\left(\phi_C^-/2\right)=-\frac{g}{2}\,\cot\left(\frac{\pi}{2}+
\delta\right).
\end{equation}                                                                  
This shows that two surface states may occur in the same gap, and in cases
where two such surface states are observed their energy separation
is a measure of the surface corrugation potential, as pointed out by
Smith \cite{smith85}.

By first introducing the experimentally determined band gap parameter 
$V_{\bf g}$
and crystal inner potential $V_0$ into either Eq. (\ref{phic}) or
Eqs. (\ref{phicp}) and (\ref{phicm}), and then
using the explicit energy dependence of $\phi_B$ dictated
by Eq. (\ref{phib}), Smith and collaborators\cite{smith85,garrett} solved
Eq. (\ref{bohr}) for the energies of the surface states
of a variety of face-centered cubic metals. Both crystal-induced
and image-potential induced surface state binding energies were
found to be in reasonable agreement with photoemission and inverse
photoemission measurements, and the systematics of surface-state occurrence
in bulk band gaps on different crystals were examined in an elegant way.

Instead of using the WKB expression for $\phi_B$ [Eq. (\ref{phib})], which is
only appropriate for energies that lie very near the vacuum level, 
Ortu\~no and Echenique \cite{ortuno} considered the more accurate expression
\begin{equation}\label{phibnew}
\phi_B=2\,\tan^{-1}\left[\frac{1}{\kappa\,y}\,\frac{J_0(y)\cos(\lambda\,\pi)
+N_0(y)\sin(\lambda\,\pi)}{J_1(y)\cos(\lambda\,\pi)+N_1(y)\sin(\lambda\,\pi)}
\right],
\end{equation}
as obtained by introducing into Eq. (\ref{phase}) the Wannier
approximation\cite{wannier} to the Whittaker functions, i.e.,
\begin{equation}
W_{\lambda,1/2}(y)\approx
y\left[J_1(y)\cos(\lambda\,\pi)+N_1(y)\sin(\lambda\,\pi)\right].
\end{equation}
Here, $y=\sqrt{2\,(z_{\rm im}+z_c)}$, and $J_0(y)$, $J_1(y)$, $N_0(y)$ and
$N_1(y)$ are Bessel and Neumann functions, respectively.

By using Eq. (\ref{phibnew}), Ortu\~no and Echenique \cite{ortuno} solved Eq.
(\ref{bohr}) for the energies of the surface states associated with the bulk $L$
gap projected onto the (111) surfaces and the bulk $X$ gap projected onto
the (100) surfaces of face-centered cubic Cu, Ag, and Ni. It was
demonstrated that this model correctly predicts the trends and systematics
of both crystal induced and image-potential induced surface states, and was
found that the experimental binding energies are reproduced for image-plane
positions that are closer to the crystal surface than those predicted by Lang
and Kohn\cite{lang2}.

\section{One-dimensional model potential}

At large distances from the surface, the one-dimensional potential $V(z)$
experienced by an electron has the asymptotic form exhibited in Fig. 3, with the
image plane located at $z=z_{\rm im}$. However, on approaching the surface
$V(z)$ must merge continuously into the bulk crystal potential. Jones,
Jennings, and Jepsen proposed a saturated model potential of the
form \cite{jjj}

\begin{equation}\label{vz}
V(z)=\left\{\begin{array}{ll}
\displaystyle{\frac{V_0}{1+A\,{\rm e}^{\alpha(z-z_{\rm im})}}} 
& \textrm{$z\le z_{\rm im}$}\\\\
\displaystyle{-\frac{1}{4\,(z-z_{\rm im})}\,\left[1-{\rm e}^{-\lambda(z-z_{\rm
im})}\right]} &
\textrm{$z_{\rm im}\le z$,}
\end{array}\right. 
\end{equation}
where $V_0$ is the inner potential, $\lambda^{-1}$ is a characteristic distance
for the changeover between the inner potential and the image asymptotic form,
and the parameters $A$ and $\alpha$ are fixed by the requirement of continuity.
By numerically integrating the Schr\"odinger equation with the model potential
of Eq. (\ref{vz}), Smith and collaborators \cite{smith89} obtained the phase
change $\phi_B$ and applied the two-band model described above to estimate
the image-plane position of a variety of metal surfaces. They found that
while the face dependence of the image-plane position is a direct result of
the discrete lattice nature of a real material, the response of the electron
density is the fundamental quantity determining the image-plane position
itself.  

Recently, a model potential
which involves the use of the $n=0$ and $n=1$
surface-state binding energies as well as the width and position of the energy 
gap as fitted parameters was proposed \cite{chulkov0}. This 
one-dimensional film model potential has the form
\begin{equation}\label{vz2}
V(z)=\left\{\begin{array}{ll}
\displaystyle{A_{10}+A_1\,\cos(2\pi z/a_s)} & \textrm{$z\le D$}\\\\
\displaystyle{A_{20}+A_2\,\cos\left[\beta\,(z-D)\right]} & 
\textrm{$D\le z\le z_1$}\\\\
\displaystyle{A_3\,{\rm e}^{-\alpha(z-z_1)}}, & \textrm{$z_1\le z\le z_{\rm im}$}\\\\
\displaystyle{-\frac{1}{4\,(z-z_{\rm im})}\,\left[1-{\rm e}^{-\lambda(z-z_{\rm
im})}\right]} &
\textrm{$z_{\rm im}\le z$},
\end{array}\right. 
\end{equation}
where the origin of coordinates has been taken to be located at the middle
plane of the film, $D$ is the half-width of the film, and $a_s$ is the interlayer
spacing. This model potential has ten parameters, $A_{20}$, $A_3$, $\alpha$,
$z_1$, $\lambda$, $z_{\rm im}$, $A_{10}$, $A_1$, $A_2$, and $\beta$, but only
four of them are independent. $A_{20}$, $A_3$, $\alpha$, $z_1$, $\lambda$, and
$z_{\rm im}$ are determined from the requirement of continuity of the potential
and its first derivative everywhere in space, and the remaining four
parameters are chosen as adjustable parameters. The parameters $A_1$ and
$A_{10}$ reproduce the width and position of the energy gap, while $A_2$ and
$\beta$ are chosen to reproduce the experimental or first-principles $n=0$ and
$n=1$ surface-state binding energies at the $\bar\Gamma$ (${\bf
k}_\parallel=0$) point. Although band-structure calculations that are based on
the model potential of Eq. (\ref{vz2}) yield work functions that are within
$0.1\,{\rm eV}$ of the measured values, for a more precise description of the
unoccupied states below the vacuum level the use of experimental work functions
is recommended.

The model potential of Eq. (\ref{vz2}) was constructed in Ref.
\cite{chulkov00} for 14 simple and noble metal surfaces. By using this
model potential both wave functions and binding energies of image states were
investigated, and the issue of the image-plane position was also addressed.
Image planes were found to be typically closer to the crystal surface than those
predicted by Lang and Kohn \cite{lang2} for a jellium surface, and they were
found to be, in the case of noble metals, in agreement with those reported in
Ref. \cite{smith89} by Smith and collaborators. Good agreement was also observed with the
image-plane positions reported in Ref. \cite{eguiluz2} from
first-principles band-structure calculations of the (111) and (100) surfaces of
Pd.

Key quantities in the description of the lifetime of surface states are
surface-state wave functions and charge densities. As shown in Ref.
\cite{chulkov00,chulkov1}, the model potential of Eq.
(\ref{vz2}) yields charge densities of both $n=0$ and $n=1$ surface states
that are in excellent agreement with those obtained with the use of
first-principles calculations.  

\section{Lifetimes}

Accurate image-state lifetimes were reported in Refs.
\cite{chulkov1,chulkov2,osma,sarria,schafer,link}, as obtained from self-consistent
many-body calculations involving the use of the one-dimensional model potential
of Eq. (\ref{vz2}).

In the framework of Green-function theory, one identifies the inverse
quasiparticle lifetime as follows
\begin{equation}\label{lifetime}
\tau_s^{-1}=-2\,{\rm Im}E_s,
\end{equation}
where $E_s$ is the quasiparticle energy. On the energy-shell \cite{hedin}, one
finds\cite{maziar}
\begin{equation}\label{lifetime2}
\tau_s^{-1}=-2\int d{\bf r}\int d{\bf r}'\,\Psi_s^*({\bf r})\,{\rm
Im}\Sigma({\bf r},{\bf r}';E_s)\,\Psi_s({\bf r}),
\end{equation}
where $\Psi_s({\bf r})$ and $E_s$ are taken to be the eigenfunctions and
eigenvalues of a suitable Hermitian single-particle Hamiltonian and
$\Sigma({\bf r},{\bf r}';E_s)$ is the so-called self-energy of the quasiparticle, which
is a nonlocal, energy-dependent, non-Hermitian operator accounting for all
xc effects beyond the Hartree approximation. Assuming
translational invariance in the plane of the surface, i.e., taking
$\Psi_s({\bf r})$ and $E_s$ to be of the form dictated by Eqs. (\ref{psi}) and
(\ref{energy}), Eq. (\ref{lifetime2}) yields
\begin{equation}\label{lifetime3}
\tau_s^{-1}=-2\int dz\int dz'\,\psi_s^*(z)\,{\rm
Im}\Sigma(z,z';{\bf k}_\parallel,\varepsilon_s)\,\phi_s(z),
\end{equation}
where $\psi_s(z)$ and $\varepsilon_s$ represent quasiparticle wave functions
and energies describing motion normal to the surface. In usual practice, these
wave functions and energies are taken to be either the eigenfunctions and
eigenvalues of a one-dimensional LDA Kohn-Sham Hamiltonian of DFT or the
solutions of a one-dimensional Schr\"odinger equation with the model potential of
Eq. (\ref{vz2}).

\subsection{$GW$ approximation}

The exact self-energy $\Sigma(z,z';{\bf k}_\parallel,\varepsilon_s)$ can be
obtained, in principle, from an iterative solution of Hedin's equations
\cite{hedin} in combination with the Dyson equation \cite{fetter}. However, to
obtain explicit results one usually resorts to an expansion in powers of the
time-ordered screened interaction $W(z,z';{\bf k}_\parallel,\varepsilon)$. The
leading order term of this expansion is the so-called $GW$ approximation:
\begin{eqnarray}\label{gw}
\Sigma(z,z';{\bf k}_\parallel,\varepsilon_s)=\int_{-\infty}^\infty
{d\varepsilon\over 2\pi}&&e^{-{\rm i}\eta\varepsilon}\,
G(z,z';{\bf k}_\parallel,\varepsilon_s-\varepsilon)\nonumber\\
&&\times\,W(z,z';{\bf
k}_\parallel\varepsilon),
\end{eqnarray}
which can also be obtained as the first iteration of Hedin's equations by
simply neglecting vertex corrections. $G(z,z';{\bf
k}_\parallel,\varepsilon)$ represents the one-particle Green
function, $\eta$ is a positive infinitesimal, and the screened interaction can
be expressed in terms of the density-response function
$\chi(z,z';{\bf k}_\parallel,\varepsilon)$, as follows
\begin{eqnarray}\label{w}
&&W(z,z';{\bf k}_\parallel,\varepsilon)=v(z-z',{\bf k}_\parallel)
+\int{\rm
d}{z_1}\int{\rm d}{z_2}\nonumber\\
&&\times v(z-z_1,{\bf
k}_\parallel)\,
\chi(z_1,z_2;{\bf k}_\parallel,\varepsilon)\,v(z_2-z').
\end{eqnarray}

Most current $GW$ calculations simply replace the {\it exact} one-particle
Green function entering Eq. (\ref{gw}) by the non-interacting Green
function
\begin{equation}\label{green0}
G^0(z,z';{\bf k}_\parallel,\varepsilon)=\sum_f\frac{\psi_f^*(z)\psi_f(z)}
{\varepsilon-\varepsilon_f+{\rm i}\eta{\rm sgn}(\varepsilon-E_F)},
\end{equation}
where $\psi_f(z)$ and $\varepsilon_f$ are taken to be a complete set
of either eigenfunctions and eigenvalues of a one-dimensional LDA
Kohn-Sham Hamiltonian or solutions of a one-dimensional Schr\"odinger equation
with the model potential of Eq. (\ref{vz2}). By introducing Eq.
(\ref{green0}) into Eq. (\ref{gw}) and then Eq. (\ref{gw}) into Eq.
(\ref{lifetime3}), one finds
\begin{eqnarray}\label{b1}
\tau_s^{-1}=-2\,\sum_f&&\int{\rm d}{z}\int{\rm d}{z'}
\int{{\rm d}{\bf q}_\parallel\over(2\pi)^2}\,\phi_{i}^*(z)\,\phi_f^*(z')\nonumber\\
&&\times{\rm Im}W(z,z';{\bf q}_\parallel,\omega)\phi_f(z)\phi_{i}(z').
\end{eqnarray}

On the same level of approximation and neglecting all vertex corrections, the
screened interaction entering Eq. (\ref{b1}) is usually obtained from Eq.
(\ref{w}) with the density-response function evaluated in the random-phase
approximation (RPA):
\begin{eqnarray}\label{rpa}
&&\chi(z,z';{\bf k}_\parallel,\varepsilon)=\chi^0(z,z';{\bf
k}_\parallel,\varepsilon)+\int{\rm d}z_1\int{\rm d}z_2\nonumber\\
&\times&\chi^0(z,z';{\bf
k}_\parallel,\varepsilon)\,v(z_1-z_2;{\bf k}_\parallel,\varepsilon)\,
\chi(z_2,z';{\bf k}_\parallel,\varepsilon),
\end{eqnarray}
$\chi^0(z,z';{\bf k}_\parallel,\varepsilon)$ being the noninteracting
density-response function
\begin{eqnarray}\label{chi0}
\chi^0(z,z';{\bf k}_\parallel,\varepsilon)=-2\,{\rm i}&&\int{\rm d}\varepsilon'\,
G^0(z,z';{\bf k}_\parallel,\varepsilon')\nonumber\\
&&\times G^0(z,z';{\bf
k}_\parallel,\varepsilon+\varepsilon').
\end{eqnarray}

\subsection{$GW\Gamma$ approximation}

Short-range xc effects, which are absent in
Eqs. (\ref{b1}) and (\ref{chi0}), can be included in the framework of
the so-called $GW\Gamma$ approximation \cite{mahan1,mahan2}. In this
approximation, the electron self-energy is of the $GW$ form, i.e., it is
given by Eq. (\ref{gw}), but with an effective screened interaction  
\begin{eqnarray}\label{wnew}
&&W(z,z';{\bf k}_\parallel,\varepsilon)=v(z-z';{\bf k}_\parallel)+\int{\rm
d}z_1\int{\rm
d}z_2\nonumber\\
&&\times\left[v(z-z_1;{\bf k}_\parallel)+f_{xc}(z_1,z_2;{\bf
k}_\parallel,\varepsilon)\right]\,\nonumber\\
&&\times\chi(z_1,z_2;{\bf k}_\parallel,\varepsilon)\,v(z_2-z';{\bf k}_\parallel),
\end{eqnarray}
the time-ordered density-response function $\chi(z_1,z_2;{\bf
k}_\parallel,\varepsilon)$ now being
\begin{eqnarray}\label{alda}
&&\chi(z,z';{\bf k}_\parallel,\varepsilon)=\chi^0(z,z';{\bf
k}_\parallel,\varepsilon) +\int{\rm d}z_1\int{\rm d}z_2\nonumber\\
&&\times\chi^0(z,z';{\bf
k}_\parallel,\varepsilon)\,\left[v(z-z';{\bf k}_\parallel)+f_{xc}(z_1,z_2;{\bf
k}_\parallel,\varepsilon)\right]\,\nonumber\\
&&\times\chi(z_2,z';{\bf k}_\parallel,\varepsilon).
\end{eqnarray}
The kernel $f_{xc}(z_1,z_2;{\bf k}_\parallel,\varepsilon)$ entering Eqs.
(\ref{wnew}) and (\ref{alda}) accounts for the reduction in the
electron-electron interaction due to the existence of short-range xc 
effects associated to the excited quasiparticle and to screening electrons,
respectively. In the so-called time-dependent local density approximation
(TDLDA) \cite{tdlda} or, equivalently, adiabatic local-density approximation
(ALDA), the {\it exact} xc kernel is replaced by
\begin{equation}\label{kernellda}
f_{xc}^{ALDA}(z,z';{\bf
k}_\parallel,\varepsilon)=\left.{d^2\left[n\,\varepsilon_{xc}^{unif}(n)\right]\over
dn^2}\right|_{n=n(z)}\delta(z-z'),
\end{equation}
where $\varepsilon_{xc}^{unif}(n)$ is the xc energy per particle of a uniform
electron gas of density $n$, and $n(z)$ is the actual electron density at
point $z$.

\subsection{Effective mass}

The dispersion $E({\bf k}_\parallel)$ of both bulk and surface states has been
determined experimentally with the use of inverse photoemission
techniques at off-normal emission \cite{fauster}, showing that it is of the form
dictated by Eq. (\ref{energy}) with ${\bf k}_\parallel/2$ replaced by ${\bf
k}_\parallel/(2m)$, $m$ representing an effective mass. Surface
corrugation effects on the effective mass of image states have been found
to be negligible, and many-body effects are found to enhance the effective
mass by no more than $\approx 2\%$. Nevertheless, the effective mass of
the $n=0$ crystal induced surface state and that of unoccupied bulk states
considerably deviate from the free-electron mass, and the impact of this
deviation on the lifetime of both crystal-induced and image-potential induced
surface states is found to be important. On the one hand, there is the effect
of the decrease of the available phase space, which is easily found to scale as
$\sqrt{m_f}$, $m_f$ being the effective mass of the various available final
states. On the other hand, as the effective mass decreases, the decay from the
image state occurs, for a given energy transfer, through smaller parallel
momentum transfers, which may result in either enlarged or diminshed screened
interactions, depending on the magnitude of momentum and energy
transfers.      

\subsection{Results}

In Table 1 we compare  theoretical results with 2PPE and TR-2PPE data for 
image states at the $\overline{\Gamma}$ point on various metal surfaces.
While the calculations reported in Ref. \cite{chulkov1} were carried
out within the $GW$ approximation with use of the free-electron mass, 
realistic effective masses were introduced in the calculations reported 
in Refs.
\cite{sarria,chusima,schafer,link}. $GW\Gamma$ calculations were only
reported in Ref. \onlinecite{sarria} for the $n=1$ image-state lifetime
on the (100) and (111) surfaces of Cu. It was found that differences
between the calculations reported in Refs. \cite{chulkov1,sarria} for
these surfaces were mainly due to the impact of the actual effective
mass on the decay mechanism, and was shown that $GW$ calculations produce
decay rates that are very close to $GW\Gamma$ calculations.  
Both GW and GW$\Gamma$ calculations produce image-state lifetimes that are 
in good agreement with TR-2PPE data for Cu(100), Cu(111), Pt(111), and Pd(111).
Even for Ag surfaces, where the polarization of $d$-electrons is important for 
the description of the surface and bulk plasmons and which has not been 
included in the calculation, GW calculations lead to a reasonable value 
for the image-state lifetime. Calculations that include the effect of
$d$-electron polarization are now in progress \cite{aran}.

\begin{table}
\caption{\label{tab:table1} Linewidth (inverse lifetime) of image states, in meV.}
\begin{ruledtabular}
\begin{tabular}{lcccc}
& & 2PPE & TR2PPE & Theory \\
\hline
Cu(100) & n=1 & $28\pm6$\footnotemark[1] & $16.5\pm3/2$\footnotemark[2,3]        &
$22^e;17$\footnotemark[6] \\
        & n=2 &            &  $5.5\pm0.8/0.6$\footnotemark[2,3]    &
$5$\footnotemark[5]        \\
        & n=3 &            &  $2.20\pm0.16/0.14$\footnotemark[2,3] &
$1.8$\footnotemark[5]          \\
     \hline
Cu(111) & n=1 & $85\pm10$\footnotemark[1] & $38\pm14/9$\footnotemark[7];
$30$\footnotemark[8] & 
$38$\footnotemark[5];$29$\footnotemark[6] \\
     \hline
Ag(100) & n=1 & $21\pm4$\footnotemark[1]    & $12\pm1$\footnotemark[2,3]&
$33$\footnotemark[11] \\
        & n=2 & $3.7\pm0.4$\footnotemark[1] & $4.1\pm0.3/0.2$\footnotemark[2,3] & \\
        & n=3 &               & $1.83\pm0.08$\footnotemark[3]                &  \\
     \hline
Ag(111) & n=1 & $45\pm10$\footnotemark[1];$55$\footnotemark[8] & $22\pm10/6$
\footnotemark[13] & \\
     \hline
Au(111) & n=1 & $160\pm40$\footnotemark[1] &  & \\
     \hline
Ni(100) & n=1 & $$ &  & \\ 
     \hline
Ni(111) & n=1 & $84\pm10$\footnotemark[1] &  & \\
     \hline
Co(0001) & n=1 & $95\pm10$\footnotemark[1] &  & \\
     \hline
Fe(110) & n=1 & $130\pm30$\footnotemark[1] &  & \\
     \hline
Pt(111) & n=1 & - & $25\pm10/5$\footnotemark[14] & $29$\footnotemark[14] \\
        & n=2 & - & $11\pm1/2$\footnotemark[14]  &  $9$\footnotemark[14] \\
     \hline
Pd(111) & n=1 & $70\pm8$\footnotemark[1];$32$\footnotemark[15] &
$27$\footnotemark[15] &$30$\footnotemark[15] \\
        & n=2 & -    & -    &  $7$\footnotemark[15] \\
     \hline
Li(110) & n=1 & - & - & $37$\footnotemark[16] \\
        & n=2 & - & - & $15$\footnotemark[16] \\
\end{tabular}
\end{ruledtabular} 
\footnotetext[1]{From Ref.~\onlinecite{fauster}.}
\footnotetext[2]{From Ref.~\onlinecite{hofer}.}
\footnotetext[3]{From Ref.~\onlinecite{shumay}.}
\footnotetext[5]{From Ref.~\onlinecite{chulkov1}.}
\footnotetext[6]{From Ref.~\onlinecite{sarria}.}
\footnotetext[11]{From Ref.~\onlinecite{chusima}.}
\footnotetext[13]{From Ref.~\onlinecite{harris}.}
\footnotetext[14]{From Ref.~\onlinecite{link}.}
\footnotetext[15]{From Ref.~\onlinecite{schafer}.}
\footnotetext[16]{From Ref.~\onlinecite{chulkov2}.} 
\end{table}

\begin{acknowledgments}
The authors gratefully acknowledge partial support by Iberdrola S.A., the University of the
Basque Country, the Basque Hezkuntza, Unibertsitate
eta Ikerketa Saila, and the Spanish Ministerio de Educaci\'on y Cultura.
\end{acknowledgments}
 
\bibliography{smith2}

\end{document}